\documentclass[11pt,onecolumn]{IEEEtran}
\usepackage{epsfig,amssymb,amsmath}
\usepackage{amsmath,hhline}
\usepackage{array,arydshln}
\usepackage{dsfont}
\usepackage{array}

\newtheorem{theorem}{Theorem}

\newtheorem{lemma}{Lemma}
\newcommand{\bs}[1]{\ensuremath{\boldsymbol{#1}}}

\begin{document}

\title{Woven Graph Codes: \\ Asymptotic Performances and Examples}


\author{\authorblockN{\hspace*{-2mm}Irina E. Bocharova and  Boris
D. Kudryashov}\\
 \authorblockA{\hspace*{-2mm} Dept. of Information Systems,\\
\hspace*{-2mm} St. Petersburg Univ. of Information Technologies, Mechanics and Optics\\
\hspace*{-2mm}St. Petersburg 197101, Russia  \\
\hspace*{-2mm}Email: \{irina, boris\}@eit.lth.se} \and\\
\authorblockN{\hspace*{-2mm} Rolf Johannesson}\\
\authorblockA{Dept. of Electro and Information Technology,
Lund University\\
\hspace*{1mm}P. O. Box 118, SE-22100 Lund, Sweden\\
Email: rolf@eit.lth.se} \and \\
\authorblockN{\hspace*{-2mm} Victor V. Zyablov}\\
\authorblockA{ Inst. for Information Transm. Problems,
 Russian Academy of Sciences\\
 Moscow 101447, Russia\\
 Email: zyablov@iitp.ru} }

\maketitle

\begin{abstract}

Constructions  of woven graph codes based on constituent block and convolutional
codes are  studied. It is shown that within  the random ensemble of such
codes based on $s$-partite, $s$-uniform hypergraphs,  where  $s$ depends only on
the code rate,   there
exist codes satisfying the Varshamov-Gilbert (VG) and  the Costello lower
bound on the minimum distance and the free distance, respectively.
A  connection between regular bipartite graphs and tailbiting
codes is shown. Some examples of woven graph codes are presented.  Among
them an example of a rate $R_{\rm wg}=1/3$ woven
graph code with  $d_{\rm free}=32$ based on Heawood's bipartite graph and containing  $n=7$
constituent   rate $R^{c}=2/3$ convolutional codes with overall
constraint lengths $\nu^{c}=5$ is given. An
encoding procedure for  woven graph codes  with complexity proportional to
the number of constituent codes and their overall constraint length
$\nu^{c}$ is presented.
\end{abstract}


\section{Introduction}

Woven graph codes can be considered as a generalization of low-density parity-check
(LDPC) block codes \cite{Gallag}. Their structure as  graph codes
makes them   suitable for iterative decoding. Moreover, the LDPC block
codes are known as codes with low-complexity decoding and
they can be considered as competitors to  the turbo codes \cite{beru}
which are sometimes called parallel concatenated codes. As mentioned in
\cite{barg},  the underlying graph defines a permutation of the
information symbols which resembles  the interleaving in turbo coding
schemes.

On the other hand, similarly to the LDPC codes, graph codes usually have minimum
distances essentially smaller than those of the best known linear
codes of the same parameters. At a  first glance,  the minimum distance of a
graph code does not play an important role in iterative decoding
since the error-correcting capability of this suboptimal procedure is
often  less than
that guaranteed by the minimum distance. However, in general, the
belief-propagation decoding algorithms work better if the {\em girth} of
the underlying graph is large, that is, if the minimum distance of the
graph code is large \cite{foss}.

In the sequel we distinguish between graph,  graph-based, and  woven graph
codes. We say that a {\em graph code} is a   block code whose parity-check matrix
coincides with the incidence matrix of the corresponding graph. {\em Graph-based
codes} constitute  a class of concatenated codes with constituent block codes
concatenated with a graph code (see, for example,
\cite{barg}). Each  vertex  in the underlying  graph corresponds to a constituent block
code. The main feature of these codes is that the block length of their
constituent block codes  coincides with the  degree of the underlying graph.

We introduce  {\em woven
graph codes} which are, in fact, graph-based codes with constituent
block codes whose  block
length is  a multiple of  the graph degree $c$, that is, their block length is  $lc$, where
$l$ is an integer. In particular, when $l$ tends to infinity we
obtain  convolutional constituent codes.

 Distance properties of bipartite graph-based codes with
constituent block codes  were  studied in \cite{barg}. It was  shown
that if the minimum distance of the constituent block codes  is
larger than or equal to $3$, then there exist asymptotically good
codes with fixed constituent codes among these graph-based codes. Also it was
shown in \cite{barg}  that for some range of rates, random graph-based  codes
with block constituent codes  satisfy the VG bound when the block
length of the constituent codes  tends
to infinity. One
disadvantage of graph-based codes  that  becomes
apparent in the asymptotic
analysis is that good performances can only be achieved when the block
length of the
constituent block codes  (which in this case coincides with the graph
degree $c$) tends
 to infinity. In practice this  leads to rather
long graph-based codes with not only rather high decoding complexity of the
iterative decoding procedures but also high encoding complexity.

In this paper, we consider a class of the generalized graph-based
codes which we call  woven  graph  codes with constituent
block and convolutional codes. They are based on {\em $s$-partite, $s$-uniform
  hypergraphs}. Notice that graph-based  codes with  constituent block
codes based on hypergraphs were considered in
\cite{zyablov}, \cite{shlomo}. It is mentioned in \cite{zyablov} that
Gallager's LDPC codes
are graph codes over hypergraphs.

  We consider first  woven graph codes with constituent
$(lc,lb)$ block  codes. A product-type lower bound on the
minimum distance of such codes is derived.
In order to analyze their asymptotic performances
we modify the approach used in \cite{barg} to  $s$-partite,  $s$-uniform hypergraphs and
constituent $(lc,lb)$ block codes\footnote{When we were preparing this paper we were
informed that the possibility of achieving the VG bound by
considering hypergraphs was known to
  A. Barg \cite{barg2}.}.    It is
shown that when $l$ grows to infinity  in the random
ensemble of  woven
graph codes with  binary constituent block codes we can
find  $s \ge 2$  such that there exist
codes satisfying  the VG lower bound on the minimum distance for any rate.

In order to generalize the asymptotic analysis to  woven graph
codes with constituent convolutional  codes we assume that the binary
constituent block code is chosen as a zero-tail (ZT) terminated
convolutional code and  consider  a sequence of ZT convolutional codes
of increasing block length $l$.  It is shown that  when the overall constraint
length of the woven graph code tends to infinity in the random
ensemble of such convolutional codes  we can find  $s\ge 2$ such that
there exist codes satisfying the Costello lower bound
on the free distance for any rate.

 We also describe the  constituent convolutional codes
 as block codes over the  field  of binary
Laurent series \cite{RolfKam}. This description as well as the notion of {\em block}
Hamming distance \cite{SidBos} of convolutional codes  is used
to derive  a product-type lower bound on the free distance of woven graph codes with
constituent convolutional codes and to construct   examples  of such
woven codes with rate $R_{\rm wg}=1/3$. For a given hypergraph the
free  distance of the woven graph code depends on the numbering of
 code symbols associating to  the hypergraph vertices. By a
 search over all possible permutations of the constituent code we
 found an example of  a  rate $R_{\rm wg}=1/3$ woven graph code with
 overall constraint length $\nu=64$ and free distance
$d_{\rm free}=32$. The rate $R_{\rm wg}=1/3$ woven graph code  is based on Heawood's bipartite graph
\cite{bondy}, \cite{harary} and contains  constituent convolutional codes with
overall constraint length $\nu^{c}=5$ and free distance $d_{\rm free}^{c}=6$.

We consider also the encoding problem for  graph and woven graph codes. The
traditional encoding technique for graph  codes has
complexity $O(N^{2})$, where $N$ is the blocklength. We show by examples
that some regular block graph codes are quasi-cyclic and thereby
can be interpreted as tailbiting (TB)  codes (see, for
example, \cite{solom}, \cite{Ma}). It is
known that the encoding complexity of such codes is proportional to the
overall constraint length of the parent convolutional code.

By using a TB representation for  the graph code  we can construct  an
example of an encoder for a woven graph code that is also represented in the form of
a TB code but with overall constraint length less than or equal to
$2n\nu^{c}$, where $n$ is the number of constituent convolutional codes
with overall constraint length $\nu^{c}$ each.

In Section II,  we  consider some properties of $s$-partite,
$s$-uniform, $c$-regular hypergraphs. We
define woven graph codes with  constituent block codes as well as with
 constituent convolutional codes and obtain product-type lower bounds
on their minimum and free distances.
Then, in Section III, we derive a lower bound on the free
distance of the random ensemble of woven graph codes. In Section IV,
examples  of  woven  graph codes are given. We conclude the paper
by considering encoding techniques for graph codes and
woven graph codes in Section V.

\section{Preliminaries}

A {\em hypergraph} is a generalization of a graph in which the  edges
are subsets of vertices and may connect (contain)  any
number of vertices.  These edges are called hyperedges. A hypergraph is called $s$-uniform if every hyperedge
has cardinality $s$ or, in other words,  connects $s$
vertices. If $s=2$ the hypergraph is simply  a graph. The {\em degree  of
a vertex}
 in a hypergraph is the number of
hyperedges that are connected to  (contain) it. If all vertices have the same degree we say
that this is the {\em degree of the hypergraph}.
 The hypergraph is {\em $c$-regular} if every vertex
has the same degree $c$.

Let the set $V$ of vertices of an  $s$-uniform hypergraph
be partitioned  into $t$ disjoint subsets $V_{j}$, $j=1,2,\dots,t$.
 A hypergraph is said to be {\em $t$-partite} if
no edge contains two vertices from the same set $V_{j}$, $j=1,2,\dots,t$.

 In the sequel we consider $s$-partite, $s$-uniform, $c$-regular
hypergraphs. Such a  hypergraph  is a
union of $s$ disjoint subsets of
vertices. Each vertex has no connections in its own set and is
connected with  $s-1$   vertices in
the other subsets.
\begin{figure}[h!]
\center{ \epsfig{file=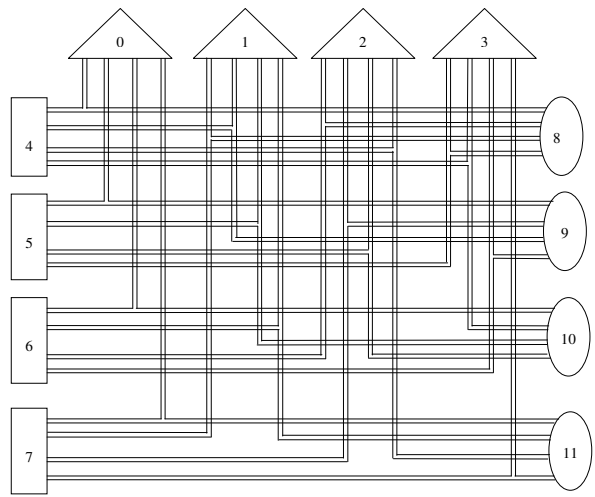,width=80mm}} \caption{A 3-partite,
  3-uniform, 4-regular hypergraph. } \label{figure:graph4}
\end{figure}
In Fig. \ref{figure:graph4} a $3$-partite, $3$-uniform, $4$-regular
hypergraph is shown. It contains three sets of vertices. They are  shown by
triangles, rectangles, and ovals, respectively. There are no edges
connecting vertices inside any of these three sets. The vertices are
connected by hyperedges each of which connects three vertices.

A {\em  cycle} of length $L$ in the hypergraph is an alternating sequence of  $L+1$
 vertices and $L$ hyperedges where all
  vertices are distinct except the initial and the final vertex, which
  coincide, and
 all edges are distinct. The
 {\em girth} of a hypergraph is the length of its shortest cycle.
\begin{figure}[h!]
\center{ \epsfig{file=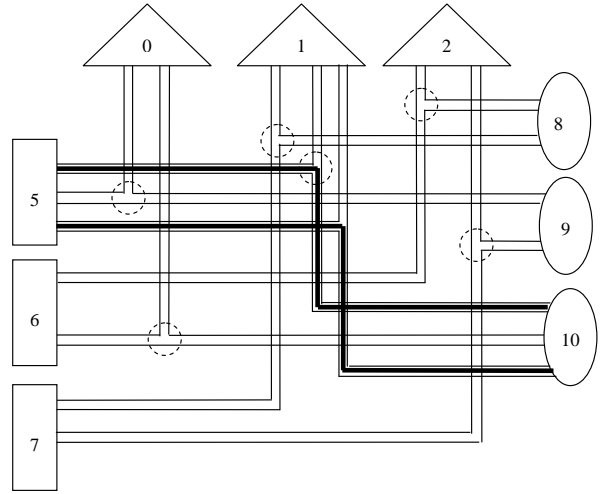,width=80mm}} \caption{A shortest
  compact subgraph. } \label{figure:hypergirth}
\end{figure}
In Fig. \ref{figure:hypergirth}  we show a subgraph that contains the shortest cycle of the $3$-partite,
$3$-uniform, $4$-regular hypergraph in Fig. \ref{figure:graph4}. It consists
of the   vertices $5$, $10$, and
$5$ and has girth equal to 2.
We introduce the notion of a {\em compact $(\ge d)$-connected subgraph} in the hypergraph. It is a
connected subgraph in which each vertex  is incident with at least $d$
hyperedges. We call  the length (number of hyperedges)
of the shortest compact subgraph its {\em ($s$,$d$)-girth}. In Fig. \ref{figure:hypergirth} the
hyperedges belonging to the shortest $(\ge 2)$-compact subgraph are marked by
circles. It is easy to see that ($3$,$2$)-girth is $6$.

\begin{figure}[h!]
\center{ \epsfig{file=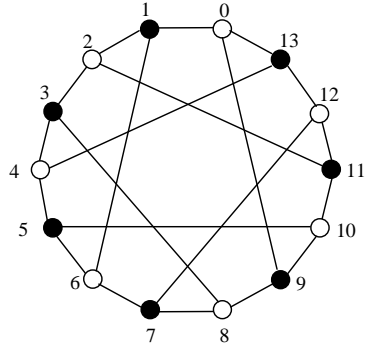,width=50mm}} \caption{Heawood's
bipartite graph. } \label{figure:graphH}
\end{figure}

 A $2$-partite, $2$-uniform hypergraph is  a bipartite graph. For such
 a hypergraph  the ($2$,$2$)-girth  is equal to the girth and a compact subgraph is
 a cycle.
Heawood's bipartite graph \cite{bondy}, \cite{harary} with $14$
vertices and $21$ edges is shown in Fig. \ref{figure:graphH}. This
graph contains a set of $n=7$ black and a set of $n=7$ white
vertices. Each vertex has no connections within its  own set
and is  connected with $c=3$ vertices from the other set. The girth of the
Heawood graph is $6$.

\subsection{Graph-based codes  and graph codes}
\begin{figure}[h!]
\center{ \epsfig{file=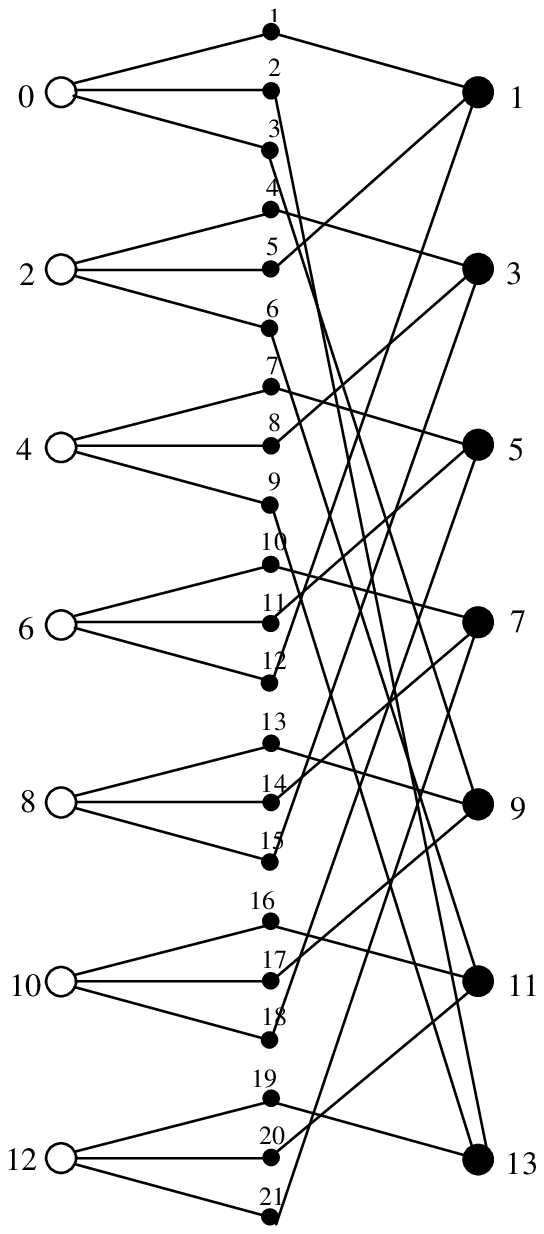,width=42mm}} %
\caption{A Tanner graph ( $c=3$, $n=7$) representation of Heawood's
bipartite graph.}
\label{figure:graph}
\end{figure}

 In order
to illustrate the structure of a binary  graph-based block  code  with
constituent block  codes  we represent the
Heawood  bipartite
graph using a so-called Tanner graph \cite{tanner81} as shown in Fig. \ref{figure:graph}.

We introduce a set of $nc=21$ (variable) vertices which correspond to the
code symbols.  Each of the $2n=14$ (constraint) vertices  on the right- and
left-hand sides corresponds to one of  14 parity checks.
The  $c=3$ edges leaving one
constraint vertex correspond to a codeword of the constituent $(c,b)$ block
code of rate $R^{c}=b/c$. The parity-check matrix of the corresponding
graph-based  code with binary constituent block codes  is
\begin{equation}
H_{\rm gb}=\begin{pmatrix}
H_{1}   \\
H_{2}
\end{pmatrix}
\label{gen2}
\end{equation}
where the parity-check matrix $H_{1}$ of size $n\times nc=7\times 21$ has the form
\[
H_{1}=\left(\begin{array}{ccccccc}
H^{c}&{\bs 0}&{\bs 0}&{\bs 0}&{\bs 0}&{\bs 0}&{\bs 0}   \\
{\bs 0}& H^{c}&{\bs 0}&{\bs 0}&{\bs 0}&{\bs 0}&{\bs 0}\\
\vdots& \vdots& \vdots&\vdots&\vdots&\vdots&\vdots\\
{\bs 0}&{\bs 0}&{\bs 0}&{\bs 0}&{\bs 0}&{\bs 0}&H^{c}
\end{array}
\right)
\]
where $H^{c}$ is a size $(c-b)\times c=(3-b)\times 3$ parity-check matrix of the
constituent block  code, and $H_{2}$ is a  size $n\times nc=7\times 21$
parity-check matrix which is the permutation of the columns of $H_{1}$
determined by the graph. Notice that in general by choosing $b<c$ and  assigning
constituent block  codes of different rates $R^{c}=b/c$ to  the same graph
we can obtain graph-based codes of different rates. In general,
since in an  $s$-partite, $s$-uniform, $c$-regular hypergraph the total number of parity checks
is equal to $sn(c-b)$,  the code rate $R_{\rm gb}$ of the
graph-based  code is
\begin{equation}
R_{\rm gb} \ge \frac{n(c-s(c-b))}{nc}=s(R^{c}-1)+1
\label{rate}
\end{equation}
 with equality if and
only if all parity-checks are
linearly independent. If $s=2$, then we get $R_{\rm gb} \ge 2R^{c}-1$.

The simplest example of a  Heawood  graph-based code   can be
obtained by choosing as  constituent block  codes  a single-parity-check
code of rate $R^{c}=1/3$. Then the parity-check matrix $H^{c}$ has the form
\[
H^{c}=
\left(
\begin{array}{lll}
1&1&1
\end{array}
\right
)
\]
and the parity-check matrix of the graph-based code is
\begin{equation}
\footnotesize
H_{\rm gb}=H_{\rm g}=
\left(
\begin{array}{ccc:ccc:ccc:ccc:ccc:ccc:ccc|r}
1&2&3&4&5&6&7&8&9&10&11&12&13&14&15&16&17&18&19&20&21 & \\ \hline
1&1&1&0&0&0&0&0&0&0&0&0&0&0&0&0&0&0&0&0&0&0 \\
0&0&0&1&1&1&0&0&0&0&0&0&0&0&0&0&0&0&0&0&0&2 \\
0&0&0&0&0&0&1&1&1&0&0&0&0&0&0&0&0&0&0&0&0&4 \\
0&0&0&0&0&0&0&0&0&1&1&1&0&0&0&0&0&0&0&0&0&6\\
0&0&0&0&0&0&0&0&0&0&0&0&1&1&1&0&0&0&0&0&0&8\\
0&0&0&0&0&0&0&0&0&0&0&0&0&0&0&1&1&1&0&0&0&10\\
0&0&0&0&0&0&0&0&0&0&0&0&0&0&0&0&0&0&1&1&1&12\\
\hdashline
1&0&0&0&1&0&0&0&0&0&0&1&0&0&0&0&0&0&0&0&0&1\\
0&0&0&1&0&0&0&1&0&0&0&0&0&0&1&0&0&0&0&0&0&3\\
0&0&0&0&0&0&1&0&0&0&1&0&0&0&0&0&0&1&0&0&0&5\\
0&0&0&0&0&0&0&0&0&1&0&0&0&1&0&0&0&0&0&0&1&7\\
0&0&1&0&0&0&0&0&0&0&0&0&1&0&0&0&1&0&0&0&0&9\\
0&0&0&0&0&1&0&0&0&0&0&0&0&0&0&1&0&0&0&1&0&11\\
0&1&0&0&0&0&0&0&1&0&0&0&0&0&0&0&0&0&1&0&0&13
\end{array}
\right)
\label{incdmatr}.
\end{equation}
In this case the graph-based code coincides with the graph code since
(\ref{incdmatr}) is the  incidence matrix of the Heawood graph.
In \cite{foss} it is proved that the minimum distance of the
bipartite graph-based code   with   single-parity-check
constituent codes is  $d_{\min}=g$, where $g$ is the girth of the
corresponding graph. Notice that for the Tanner graph we have
$d_{\min}=g/2$.
The parity-check matrix (\ref{incdmatr}) is a
$14 \times 21$ parity-check matrix. Taking into account
that one check is linearly dependent on the other,  we obtain a $(21,8)$ binary block code. Its
minimum distance is  $d_{\min}=g=6$.

Consider the hypergraph shown in Fig. \ref{figure:graph4}. Its
incidence matrix has the form
\begin{equation}
\footnotesize
H_{\rm hg}=H_{\rm hgb}=
\left(
\begin{array}{cccc:cccc:cccc:cccc|r}
 1&2&3&4&5&6&7&8&9&10&11&12&13&14&15&16&\\ \hline
 1&1&1&1&0&0&0&0&0&0&0&0&0&0&0&0&0\\
 0&0&0&0&1&1&1&1&0&0&0&0&0&0&0&0&1\\
 0&0&0&0&0&0&0&0&1&1&1&1&0&0&0&0&2\\
 0&0&0&0&0&0&0&0&0&0&0&0&1&1&1&1&3\\
 \hdashline
 1&0&0&0&0&1&0&0&0&0&1&0&0&0&0&1&4\\
 0&1&0&0&0&0&1&0&0&0&0&1&1&0&0&0&5\\
 0&0&1&0&0&0&0&1&1&0&0&0&0&1&0&0&6\\
 0&0&0&1&1&0&0&0&0&1&0&0&0&0&1&0&7\\
 \hdashline
 1&0&0&0&1&0&0&0&1&0&0&0&1&0&0&0&8\\
 0&1&0&0&0&1&0&0&0&1&0&0&0&1&0&0&9\\
 0&0&1&0&0&0&1&0&0&0&0&1&0&0&0&1&10\\
 0&0&0&1&0&0&0&1&0&0&1&0&0&0&1&0&11\\
\end{array}
\right)
\label{hyper}
\end{equation}
and  is a $12\times 16$ parity-check matrix of a hypergraph-based code
which coincides with the parity-check matrix of the hypergraph code. Each column
represents  a hyperedge  and each row represents
a vertex of this hypergraph. For example, the first four rows
represent the vertices 1, 2, 3, and 4 (triangles), the next four rows
he vertices 5, 6, 7, and 8 (rectangles), and the last four rows the
vertices 9, 10, 11, and 12 (ovals). The first column represents the
hyperedge which connects the vertices 1, 5, and 9, the second column
the hyperedge connecting vertices 1, 6, and 10 {\em etc}. The rows of (\ref{hyper}) are linearly
dependent. By removing two  parity checks we obtain a
$(16,6)$ linear block code with the minimum distance
$d_{\min}=g_{3,2}=6$, where $g_{3,2}$ is the ($3$,$2$)-girth of the hypergraph.
The rate of this hypergraph code is $R_{\rm hg}=3/8$, which satisfies inequality (\ref{rate}),

\[R_{\rm hg} \ge 3\left(\frac{3}{4}-1\right)+1=\frac{1}{4}.\]

\begin{figure}[h!]
\center{ \epsfig{file=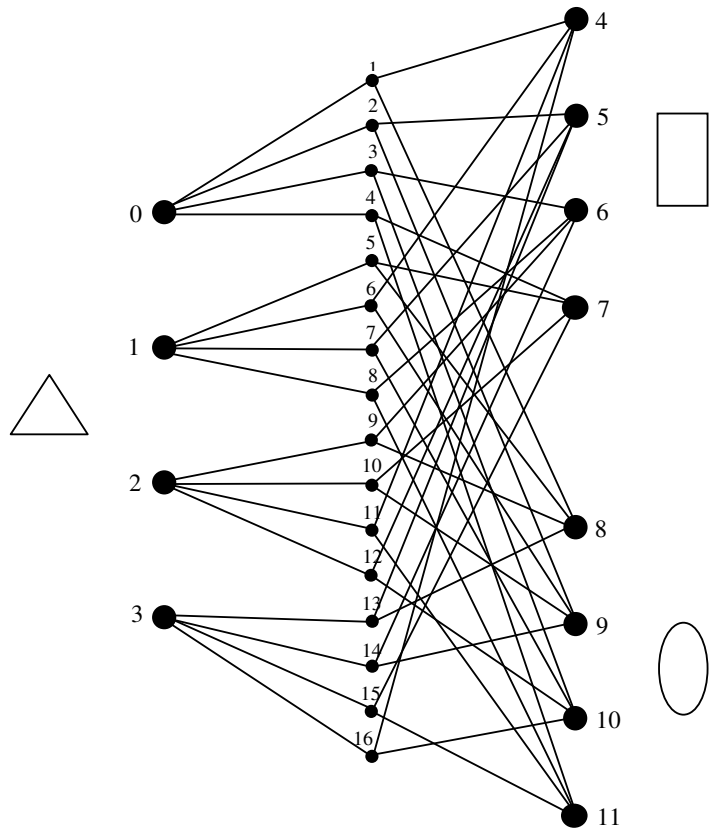,width=100mm}} %
\caption{A Tanner graph representation of the $(16,6)$ hypergraph-based code.}
\label{figure:TannerHyperGraph}
\end{figure}

The Tanner version of this hypergraph is shown in Fig. \ref{figure:TannerHyperGraph}.

For an $s$-partite, $s$-uniform, $c$-regular hypergraph-based code
with  constituent block codes we have  the following
theorem.
\begin{theorem}
\label{graph-based}
The minimum distance of   a hypergraph-based code based on an   $s$-partite,
$s$-uniform,  $c$-regular hypergraph with  ($s$,$d_{\min}^{c}$)-girth $g_{s,d_{\min}^{c}}$
and containing
constituent block codes with   minimum distance $d_{\min}^{c} \ge 2$ is

 \[d_{\min} = g_{s,d_{\min}^{c}}.\]

\end{theorem}
\vspace{5mm}
{\em Proof}. Any  nonzero
codeword in an  $s$-partite, $s$-uniform, $c$-regular hypergraph-based
code always corresponds to   a connected $(\ge d_{\min}^{c})$-subgraph or a set of disjoint
connected
subgraphs.  These subgraphs are called {\em active} \cite{tan2001}, \cite{foss}.
All hyperedges and vertices in an active subgraph are
also called {\em active}.  The number of hyperedges in the shortest connected subgraph
is equal to $g_{s,d_{\min}^{c}}$.
   Any nonzero symbol in a codeword corresponds  to an active
hyperedge in the graph.
By  using the arguments given above,   we conclude that for any codeword ${\bs v}$,
\[  w_{\rm H}({\bs v})\ge g_{s,d_{\min}^{c}}\]
where $w_{\rm H}({\bs v})$ is the Hamming  weight of ${\bs
  v}$. Minimizing over ${\bs v}$ completes
the proof.
\subsection{Woven graph codes with constituent block codes }
Now assume that the constituent  code  assigned to  the hypergraph vertices is
a binary $(lc,lb)$ linear block code determined by a  parity-check matrix
\begin{equation}
H^{c}=
\left(
\begin{array}{llll}
H^{c}_{11}& H^{c}_{12}& \dots& H^{c}_{1,c}\\
H^{c}_{21}& H^{c}_{22}& \dots& H^{c}_{2,c}\\
\vdots&\vdots&\ddots& \vdots\\
H^{c}_{(c-b),1}& H^{c}_{(c-b),2}& \dots& H^{c}_{(c-b),c}
\end{array}
\right)
\label{nonbin}
 \end{equation}
where $H^{c}_{ij} \in {\mathcal B}_{l \times l}$ is a size $l\times l$  matrix,
${\mathcal B}_{l \times l}$ is  the set of all possible binary matrices of  size $l\times l$.

Let
${\mathcal C}_{2}(H^{c})$ denote such a  binary $(lc,lb)$ constituent block code determined
by the matrix (\ref{nonbin}). We call the corresponding hypergraph-based code with
${\mathcal C}_{2}(H^{c})$ as constituent  codes a  {\em
woven graph} code with constituent block codes.

\begin{figure}[h!]
\center{ \epsfig{file=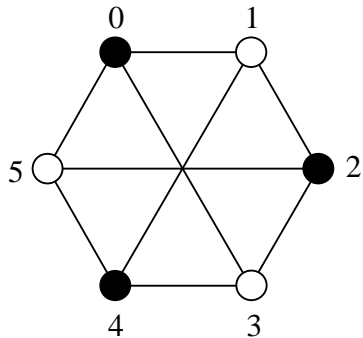,width=50mm}} \caption{Utility bipartite
graph. } \label{figure:bgraph}
\end{figure}

\begin{figure}[h!]
\center{ \epsfig{file=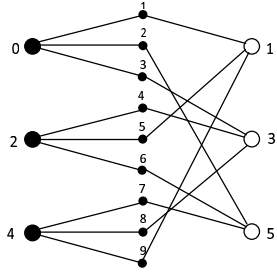,width=42mm}} %
\caption{A Tanner graph  representation of the utility bipartite graph.}
\label{figure:Tgraph}
\end{figure}

Consider an example of a woven graph code based on the bipartite graph
with girth $g=4$
shown in Fig. \ref{figure:bgraph}. The Tanner version of this the so-called
``utility''  bipartite graph
is shown  in Fig. \ref{figure:Tgraph}.

The incidence matrix of this graph is
\begin{equation}
H_{\rm g}=
\left(
\begin{array}{ccc:ccc:ccc}
1&1&1&0&0&0&0&0&0\\
0&0&0&1&1&1&0&0&0\\
0&0&0&0&0&0&1&1&1\\
\hdashline
1&0&0&0&1&0&0&0&1\\
0&0&1&1&0&0&0&1&0\\
0&1&0&0&0&1&1&0&0
\end{array}
\right).
\end{equation}
We use  a constituent ($4\times 3, 4\times 2$) linear block code with
$d_{\min}^{c}=3$
determined by the  parity-check matrix
 \begin{eqnarray*}
H^c&=&\left( H_{1}^{c}\quad H_{2}^{c}\quad H_{3}^{c} \right) \\
 &=& \left(
 \begin{array}{cccc:cccc:cccc}
 1&0&0&0&  1&1&1&0&  1&1&0&0 \\
 0&1&0&0&  0&1&1&1&  0&1&1&0 \\
 0&0&1&0&  1&0&1&1&  0&0&1&1 \\
 0&0&0&1&  1&1&0&1&  1&0&0&1 \\
 \end{array}
 \right).
\end{eqnarray*}
By searching over all possible permutations of the matrices $H_{1}^{c}$,
$H_{2}^{c}$, and $H_{3}^{c}$ we found the following parity-check
matrix of the woven graph code with the best minimum distance

\begin{equation}
\footnotesize
H_{\rm wg}= \left(
\begin{array}{ccc:ccc:ccc}
H_{1}^{c}    &H_{2}^{c}    &H_{3}^{c}    &\bf 0&\bf 0&\bf 0&\bf 0&\bf 0&\bf 0 \\
\bf 0&\bf 0&\bf 0&H_{1}^{c}    &H_{2}^{c}    &H_{3}^{c}    &\bf 0&\bf 0&\bf 0 \\
\bf 0&\bf 0&\bf 0&\bf 0&\bf 0&\bf 0&H_{1}^{c}    &H_{2}^{c}    &H_{3}^{c}     \\
\hdashline
H_{2}^{c}  &\bf 0&\bf 0&\bf 0&H_{3}^{c}    &\bf 0&\bf 0&\bf 0&H_{1}^{c} \\
\bf 0&\bf 0& H_{1}^{c}   &H_{2}^{c}    &\bf 0&\bf 0&\bf 0&H_{3}^{c} &\bf 0\\
\bf 0&H_{3}^{c}    &\bf 0&\bf 0&\bf 0&H_{1}^{c}    &H_{2}^{c}    &\bf 0&\bf 0\\
\end{array}
\right).
\label{wovbl}
\end{equation}
The matrix (\ref{wovbl}) describes a  $(36,12)$ linear block code
with $d_{\min}=10$.

Any  codeword ${\bs v}^{c}$ of the $(lc,lb)$ constituent block code can be represented as
a sequence of $c$ blocks of length $l$, that is,  ${\bs v^{c}}=({\bs
  v}_{1}^{c},{\bs v}_{2}^{c}, \dots, {\bs v}^{c}_{c})$, where ${\bs
    v}_{i}^{c}=(v^{c}_{i1},v^{c}_{i2}, \dots, v^{c}_{il})$, $i=1,2,\dots,c$.
We define the \emph{minimum Hamming block distance} between the
codewords ${\bs v}^{c}$  and ${\bs{\tilde  v}}^{c}$ of the constituent block code as
\[d_{block}^{c}=\min_{ {\bs v}^{c} \ne {\bs
   {\tilde  v}}^{c}}\left\{w_{block}({\bs v}^{c}-{\bs {\tilde v}}^{c})\right
\}\]
where $w_{block}^{c}({\bs v}^{c})=\#({\bs v}_{i}^{c}\ne {\bs 0})$,
$i=1,2,\dots,c$.
Next  we will prove the following theorem.
\begin{theorem}
\label{block}
The minimum distance of woven graph codes  based on $s$-partite,
$s$-uniform,  $c$-regular hypergraphs
with ($s$,$d_{block}^{c}$)-girth $g_{s,d_{block}^{c}}$
and containing  constituent block codes with  minimum distance
$d_{\min}^{c}$ and  minimum block distance $d_{block}^{c}\ge 2$ can be
lower-bounded by
\[d_{\min}\ge \max\left \{\frac{g_{s,d_{block}^{c}}}{c},s\right\}d_{\min}^{c}.\]

\end{theorem}

{\em Proof}. Any nonzero codeword corresponds to an active connected
subgraph or a set of disjoint connected subgraphs and the number of
hyperedges in the shortest subgraph is $g_{s,d_{block}^{c}}$.
 Any nonzero symbol in a codeword activates a hyperedge in the
 graph, that is, not less than $s$ constituent subcodes correspond to
 a codeword. Since at most $c$ hyperedges are connected with any hypergraph
 vertex then  the number of
active constituent subcodes   can  be lower-bounded by
\[ \frac{g_{s,d_{block}^{c}}}{c}.\]
Taking into account  that  any  codeword
of block weight greater than or equal to $d_{block}^{c}$ in the
constituent block  code has  a weight at least equal to $d_{\min}^{c}$
we obtain the following inequality
\[w_{\rm H}({\bs v})\ge\max\left\{
\frac{g_{s,d_{block}^{c}}}{c},s\right\}d_{\min}^{c}\]
for any codeword ${\bs v}$ and the proof is complete.

From Theorem \ref{block} for the woven graph code determined by
(\ref{wovbl}) we obtain that $d_{\min} \ge \max\left\{\frac{4}{4},3\right\}3=9.$

\subsection{Woven graph codes with constituent convolutional  codes}

Woven  graph codes  with constituent convolutional  codes can be considered as a
straightforward generalization of the  woven graph code with
constituent block  codes.  Assume that the ${\mathcal C}_{2}(H^{c})$ code is chosen as a
zero-tail terminated (ZT) convolutional code and consider a sequence
of ZT convolutional codes with increasing  $l$. It is evident that
when  $l$ tends to infinity  the  $(lc,lb)$ constituent
code ${\mathcal C}_{2}(H^{c})$ can be chosen  as a rate $R^{c}=b/c$
binary convolutional code with  constraint length $\nu^{c}$. Then the
corresponding woven  graph code has rate $R=s(R^{c}-1)+1$ and its
constraint length is at most $sn\nu^{c}$.

Another description of  woven  graph codes  with constituent convolutional
codes follows from
the representation of the constituent  convolutional code in polynomial form.
Let $G^{c}(D)$ be a minimal encoding matrix \cite{RolfKam} of a rate $R^{c}=b/c$, memory $m^{c}$
convolutional code, given in polynomial form, that is,
\begin{equation}\vspace*{2mm}
 G^{c}(D)=\begin{pmatrix}
g^{c}_{11}(D) & \dots  & g^{c}_{1c}(D) \\
 \vdots   & \ddots & \vdots  \\
g^{c}_{b1}(D) & \dots  & g^{c}_{bc}(D)
\end{pmatrix}
\label{10} \end{equation} where
$g^{c}_{ij}(D)=g_{ij}^{c(0)}+g_{ij}^{c(1)}D+g_{ij}^{c(2)}D^{2}+\dots+g_{ij}^{c(m)}D^{m}$,
$i=1,2,\dots,b$, $j=1,2,\dots,c$,  are binary polynomials such that
$m^{c}=\max_{i,j}\{\deg g^{c}_{ij}(D)\}$. The overall constraint length is
$\nu^{c}=\sum_{i}\max_{j}\{\deg g^{c}_{ij}(D)\}$. The binary information
sequence ${\bs u}^{c}(D)=(u_{1}^{c}(D), u_{2}^{c}(D), \dots,u_{b}^{c}(D))$ is
encoded as
\[{\bs v}^{c}(D)={\bs u}^{c}(D)G^{c}(D)\] where ${\bs v}^{c}(D)=(v_{1}^{c}(D), v_{2}^{c}(D),
\dots, v_{c}^{c}(D))$ is a binary code sequence. Let $H^{c}(D)$ denote a
parity-check matrix for  the same code,
\begin{equation}
H^{c}(D)=\begin{pmatrix}
h_{11}^{c}(D)& \dots & h_{1c}^{c}(D)  \\
\vdots   & \ddots & \vdots  \\
h_{r1}^{c}(D)& \dots  & h_{rc}^{c}(D)
\end{pmatrix}
\label{18}
\end{equation}
where $r=c-b$ is the redundancy of the constituent  code.

We denote by ${\mathds
F}_{2}((D))$ the field of binary Laurent series  and
regard a rate $R^{c}=b/c$ constituent  convolutional code as a
rate $R^{c}=b/c$ block code ${\mathcal C}^{c}$ over the field of binary
Laurent series encoded by $G^{c}(D)$. Then its codewords ${\bs v}^{c}(D)$
are elements of ${\mathds F}_{2}((D))^{c}$, which is the
$c$-dimensional vector space over the field of binary Laurent
series \cite{RolfKam}.

The minimum Hamming  {\em block} distance between the
codewords ${\bs v}_{j}(D)$ and ${\bs v}_{k}(D)$  is
defined \cite{SidBos} as

\[d_{\rm block}=\min_{{\bs
    v}_{j}(D)\ne{\bs v}_{k}(D)}\{w_{\rm block}({\bs v}_{j}(D)-{\bs v}_{k}(D))\}\]

\noindent where $w_{\rm block}({\bs v}(D))=\#( v_{i}(D)\ne 0)$ is
the Hamming (block) weight of ${\bs v}(D)=(v_{1}(D),v_{2}(D),\dots,v_{c}(D))$.

Representing a  convolutional code as a block code over the
field of binary Laurent series we can obtain a woven graph
code with constituent convolutional  codes as a generalization of a graph-based
code with binary constituent block codes.
For  example, a  parity-check matrix $H_{\rm wg}(D)$ of the  rate
$R_{\rm wg}=4/3-1=1/3$
Heawood's graph-based code  with $R^{c}=2/3$ constituent
convolutional codes  has the form
 \begin{equation}
H_{\rm wg}(D)=
\footnotesize
\left(
\begin{array}{ccc:ccc:ccc:ccc:ccc:ccc:ccc}
h^{c}_{1}&h^{c}_{2}&h^{c}_{3}&0&0&0&0&0&0&0&0&0&0&0&0&0&0&0&0&0&0\\
0&0&0&h^{c}_{1}&h^{c}_{2}&h^{c}_{3}&0&0&0&0&0&0&0&0&0&0&0&0&0&0&0\\
0&0&0&0&0&0&h^{c}_{1}&h^{c}_{2}&h^{c}_{3}&0&0&0&0&0&0&0&0&0&0&0&0\\
0&0&0&0&0&0&0&0&0&h^{c}_{1}&h^{c}_{2}&h^{c}_{3}&0&0&0&0&0&0&0&0&0\\
0&0&0&0&0&0&0&0&0&0&0&0&h^{c}_{1}&h^{c}_{2}&h^{c}_{3}&0&0&0&0&0&0\\
0&0&0&0&0&0&0&0&0&0&0&0&0&0&0&h^{c}_{1}&h^{c}_{2}&h^{c}_{3}&0&0&0\\
0&0&0&0&0&0&0&0&0&0&0&0&0&0&0&0&0&0&h^{c}_{1}&h^{c}_{2}&{\bs
  h}^{c}_{3}\\
\hdashline
t^{c}_{1}&0&0&0&t^{c}_{3}&0&0&0&0&0&0&t^{c}_{2}&0&0&0&0&0&0&0&0&0\\
0&0&0&t^{c}_{1}&0&0&0&t^{c}_{3}&0&0&0&0&0&0&t^{c}_{2}&0&0&0&0&0&0\\
0&0&0&0&0&0&t^{c}_{1}&0&0&0&t^{c}_{3}&0&0&0&0&0&0&t^{c}_{2}&0&0&0\\
0&0&0&0&0&0&0&0&0&t^{c}_{1}&0&0&0&t^{c}_{3}&0&0&0&0&0&0&t^{c}_{2}\\
0&0&t^{c}_{2}&0&0&0&0&0&0&0&0&0&t^{c}_{1}&0&0&0&t^{c}_{3}&0&0&0&0\\
0&0&0&0&0&t^{c}_{2}&0&0&0&0&0&0&0&0&0&t^{c}_{1}&0&0&0&t^{c}_{3}&0\\
0&t^{c}_{3}&0&0&0&0&0&0&t^{c}_{2}&0&0&0&0&0&0&0&0&0&t^{c}_{1}&0&0
\end{array}
\right)
\label{woven}
\end{equation}
where $h^{c}_{i}$ and $t^{c}_{i}$ are short-hand for
$h^{c}_{i}(D)$ and $t^{c}_{i}(D)$, respectively, and
$H^{c}(D)=\left(h_{1}^{c}(D)\quad h_{2}^{c}(D) \quad h_{3}^{c}(D)\right)$ is a
parity-check matrix of the  rate $R^{c}=2/3$ constituent convolutional
code and $(t^{c}_{1}(D),t^{c}_{2}(D),t^{c}_{3}(D))$ is one of
six possible permutations of $h_{1}^{c}(D),h^{c}_{2}(D),h^{c}_{3}(D)$.

Exploiting  the above definitions we can interpret this bipartite
woven graph-based  code  with constituent convolutional codes as follows.
The left column of vertices  in Fig. \ref{figure:graph} represents $n$ parity checks each of
which determines one of $n$ constituent fixed and identical
convolutional codes and their $nc$ branches represent
the elements $v_{ij}^{c\mathrm{L}}(D) \in {\mathds F}_{2}((D))$,
$i$ even, $0\le i\le 2n-2$, $1\le j\le c$. Similarly, the right
column of vertices  represents same
convolutional codes and their $nc$ branches represent the elements
$v_{ij}^{c\mathrm{R}}(D) \in {\mathds F}_{2}((D))$, $i$ odd, $1\le
i\le 2n-1$, $1\le j\le c$, where the set $\{v_{ij}^{c\mathrm{R}}(D)\}$
is a random permutation of the set $\{v_{ij}^{c\mathrm{L}}(D)\}$
determined by the graph.

We can also regard the $n$ left constituent convolutional codes as a
warp with $nc$ threads. Each of the $n$ right constituent
convolutional codes are tacked  on $c$ of the threads in the warp such that
each thread of the warp is tacked on exactly once. Thus, our construction is a
special case of a woven code \cite{stefrolfvic} and  we call this
graph-based code a {\em woven graph} code.

\begin{theorem}
\label{wovbound}
The free distance of a  woven graph code based on an $s$-partite,
$s$-uniform,  $c$-regular hypergraph
with the ($s$,$d_{block}^{c}$)-girth $g_{s,d_{block}^{c}}$ and containing constituent convolutional  codes with
free distance $d_{\rm free}^{c}$ and  minimum block distance
$d_{block}\ge 2$ can be lower-bounded by
\[d_{{\rm free}}\ge\max
\left\{\frac{g_{s,d_{block}^{c}}}{c},s\right\}d_{\rm free}^{c}.\]

\end{theorem}
\vspace{5mm}
{\em Proof}. Since woven graph codes  with constituent convolutional codes can
be considered as a generalization of woven graph codes  with
constituent block
codes,  the theorem follows  from Theorem \ref{block} when $l$ tends to
infinity.

For a woven graph code based on a bipartite graph with  girth $g$  and
containing constituent convolutional codes with  minimum block
distance $d_{block}^{c}=2$ and  free distance $d_{\rm free}^{c}$  by a
straightforward generalization of the
approach of \cite{foss} we obtain the following tighter bound on the free
distance
\begin{equation}d_{{\rm free}}\ge \max\left\{\frac{g}{2},2\right\}d^{c}_{\rm
    free}.\label{bipartbound}\end{equation}

\section{Asymptotic bounds on the minimum distance of woven graph
  codes}

We will show that the ensemble of random woven graph codes based on
 random $s$-partite, $s$-uniform, $c$-regular hypergraphs with a fixed degree $c$ and
with a fixed number of vertices $n$ in each subgraph
contains asymptotically good codes. In order to
prove this we will modify the approach in \cite{barg}.
\subsection{Woven graph
  codes  with constituent block  codes}
\vspace*{4mm}
First we consider the  ensemble of  random  woven graph codes with
rate $R^{c}=b/c$ constituent block  codes  determined  by the edges of
a random $s$-partite, $s$-uniform, $c$-regular hypergraph
corresponding  to
the time-varying random parity-check matrix
\begin{equation}
H_{\rm wg}=\begin{pmatrix}
{\tilde H}_{1}   \\
{\tilde H}_{2} \\
\vdots   \\
{\tilde H}_{s}
\end{pmatrix}
=
\begin{pmatrix}
\pi_{1}(H_{1})\\
\pi_{2}(H_{2})\\
\vdots\\
\pi_{s}(H_{s})
\end{pmatrix}
\label{genmatr}
\end{equation}
where
${\tilde H}_{i}=\pi_{i}(H_{i})$, $i=1,2,\dots,s$,
is a block  matrix of size $nc(1-R^{c}) \times nc$ (or a binary matrix
of size $n(c-b)l\times ncl$)
and  $\pi_{i}$ denotes a  random permutation
of the columns of $H_{i}$,
\begin{equation}
H_{i}=\begin{pmatrix}
H_{i}^{c(1)} & \bs 0   & \dots & \bs 0  \\
\bs 0    & H_{i}^{c(2)}& \bs 0  &\dots  \\
\vdots   & \dots   & \ddots &\vdots\\
\bs 0    & \dots   & \bs 0  &H_{i}^{c(n)}
\end{pmatrix}
\label{30}
\end{equation}
where
$H_{i}^{c(t)}$, $t=1,\dots,n$, denotes the  random parity-check matrix
(\ref{nonbin}) which determines the   $(lc,lb)$ constituent  block
code and  $n$ is the number  of  constituent codes in each  subgraph.

\emph{Remark}:
In \cite{barg} a more restricted  ensemble of random codes is studied in which all
matrices   are identical random matrices. In the proof of Theorem 1  we
need that the syndrome components are  independent random variables in the
product probability space of random matrices and random permutations.
The following simple example shows that this  is not always the case if
all matrices   are identical.

Consider $n=1$ constituent block codes of  block  length $c=2$ with
 $b=1$ information symbols. This example is rather
artificial since the rate of the constituent block code $R^c=1/2$ and  therefore the
rate of the graph-based  code with $s=2$ is $R_{\rm wg}=s(R^{c}-1)+1=2R^c-1=0$. In this case the parity-check matrix of the
code has the form
\[
H_{\rm wg}=
\left(
  \begin{array}{c}
    H_1 \\
   \pi(H_2) \\
  \end{array}
\right)
\]
where $\pi$  is a  random permutation of $c$ elements.
First assume that all  matrices are identical, that is,  $H_1=H_2$.
There are only 8 equiprobable elements in the product space, namely,
\begin{eqnarray*}
 \{H_{\rm wg}\}&=&
\left\{
\left(  \begin{array}{cc}     0&0 \\   0&0 \\   \end{array}\right),
\left(  \begin{array}{cc}     0&0 \\     0&0 \\   \end{array}\right),
\left(  \begin{array}{cc}     0&1 \\     0&1 \\   \end{array}\right),
\left(  \begin{array}{cc}     0&1 \\    1&0 \\  \end{array}  \right),
  \right.\\
  &&
  \left.
\left(  \begin{array}{cc}     1&0 \\    1&0 \\  \end{array}\right),
\left(  \begin{array}{cc}    1&0 \\    0&1 \\  \end{array}\right),
\left(  \begin{array}{cc}    1&1 \\    1&1 \\  \end{array}\right),
\left(  \begin{array}{cc}    1&1 \\    1&1 \\  \end{array} \right)
\right\}.
\end{eqnarray*}
For any vector ${\bs x}$  of weight 1 we have the following set of random
equiprobable syndromes:
\begin{eqnarray*}
 \{{\bs x} H_{\rm wg}^{\rm T}\}&=&
\left\{
\left(  \begin{array}{cc}    0&0 \\  \end{array}\right),
\left(  \begin{array}{cc}    0&0 \\  \end{array}\right),
\left(  \begin{array}{cc}    0&0 \\  \end{array}\right),
\left(  \begin{array}{cc}    0&1 \\  \end{array}  \right),
  \right.\\
  &&
  \left.
\left(  \begin{array}{cc}    1&1 \\  \end{array} \right),
\left(  \begin{array}{cc}    1&0 \\  \end{array}\right),
\left(  \begin{array}{cc}    1&1 \\   \end{array}\right),
\left(  \begin{array}{cc}    1&1 \\   \end{array}\right)
\right\}.
\end{eqnarray*}
 Therefore,
 \[
 P({\bs x}H_{\rm wg}^{\rm T}={\bs 0}|w_{\rm H}({\bs x})=1)=\frac{3}{8}>\frac{1}{4}.
 \]
    If $H_1$  and $H_2$ are both random and independent this probability is
 equal to 1/4.
\vspace{0.5cm}


Although this remark contradicts  the proof of Theorem 3 in
\cite{barg}, there exists another
(combinatorial) way to prove  the same statement for identical $H_i$
\cite{bassal}.
\vspace{0.5cm}

Next we prove the following theorem.

\begin{theorem} \label{VGtheorem}
(Varshamov-Gilbert lower bound)
For any  $\epsilon > 0$, some  $l_{0}>0$, some integer $s>0$
and for all $l>l_{0}$ in the random ensemble of
length $ncl$ woven graph codes with $(lc,lb)$ binary block constituent
codes of rate $R^{c}=b/c$ there exist codes of rate $R_{\rm wg}=s(R^{c}-1)+1$ such that their relative minimum  distance $\delta_{\rm wg}=d_{\min}/ncl$ satisfies the inequalities
  \begin{equation}
\delta_{\rm wg}\ge
\left\{
\begin{array}{ll}
\delta(R_{\rm wg})-\epsilon{\mbox,} &{\mbox{ if }}R_{\rm
  wg}>1+s\log_{2}(1-\delta_{{\rm \tiny VG}}(R_{\rm wg}))\\
\delta_{{\rm \tiny VG}}(R_{\rm wg})-\epsilon{\mbox,}&  {\mbox{ if
  }}R_{\rm wg}\le 1+s\log_{2}(1-\delta_{{\rm \tiny VG}}(R_{\rm wg}))
\end{array}
\right.
\label{VGbound}
\end{equation}
\end{theorem}
where $\delta(R_{\rm wg})$ is a root of the equation
\[(1-s)h(\delta)-\delta s\log_{2}\left(2^{-(R_{\rm wg}-1)/s}-1\right)=0\] and
$\delta_{\rm VG}(R_{\rm wg})$ is the solution of $h(\delta)+R_{\rm wg}-1=0$, and $h(\cdot)$
denotes the binary entropy function.

{\em Proof.}
Let $w$
be the Hamming weight of the codeword ${\bs v}$ of the random binary woven graph
code ${\mathcal C}_{2}(H_{\rm wg})$. We are going to
find a  parameter $d$ such that  the probability ${\rm
  P}({\bs v}H_{\rm wg}^{\mathrm T}={\bs 0}|w)$ tends
to 0 for all $w < d$.
We can rewrite ${\rm P}({\bs v}H_{\rm wg}^{\mathrm T}={\bs 0}|w)$ as
\begin {equation}{\rm P}({\bs v}H_{\rm wg}^{\mathrm T}={\bs 0}|w)=\sum_{{\bs
      j}}{\rm P}({\bs v}H_{\rm wg}^{\mathrm T}={\bs 0}|w,{\bs j}){\rm P}({\bs j}|w)
\label{main}
\end{equation}
where ${\bs j}=(j_{1},j_{2},\dots,j_{s})$ and $j_{i}$ denotes the
number of nonzero constituent codewords in the $i$th subgraph
corresponding to the codeword of weight $w$.

In the ensemble of random  parity-check matrices $H^{c(t)}$,
$t=1,2,\dots,n$,  of size
$lc(1-R^{c})\times lc$   the
probability that a  nonzero vector $\bs v^{c}$ is  a codeword of the
corresponding constituent random binary code ${\mathcal C}_{2}(H^{c})$
is equal to $2^{-(c-b)l}$ since the syndromes of the constituent  codes are
equiprobable  sequences of length $(c-b)l$.
Taking into account that in the $i$th subgraph we have $j_{i}$
nonzero constituent  codewords the probability ${\rm P}({\bs
  v}H_{\rm wg}^{\mathrm T}={\bs 0}|w,{\bs j})$ can be upper-bounded by
\begin{equation}
{\rm P}({\bs v}H_{\rm wg}^{\mathrm T}={\bs 0}|w,{\bs j})\le
\binom{ncl}{w}\prod_{i=1}^{s}2^{-j_{i}cl(1-R^{c})}.
\label{redunds}
\end{equation}

In order to estimate the probability ${\rm P}({\bs j}|w)$ we prove the
following lemma.

\begin{lemma} For the ensemble of binary woven graph codes with constituent
  block codes described in
Theorem \ref{VGbound},  the probability ${\rm P}({\bs j}|w)$ that a codeword of weight $w$
contains ${\bs j}=(j_{1},j_{2},\dots,j_{s})$ nonzero constituent
codewords in the $s$ subgraphs can be upper-bounded by
\begin{equation}
P\left({\bs j}|w\right)\le \prod_{i=1}^{s}\frac{\binom{n}{j_{i}}\binom{cl}{w/j_{i}}^{j_{i}}\binom{w-1}{j_{i}-1}}{\binom{ncl}{w}}.
\label{lemf1}
\end{equation}
\label{Lemma}
\end{lemma}
{\em Proof.} Taking into account that in the $i$th  subgraph the
number of nonzero component codewords is equal to $j_{i}$ and that the
subgraphs are  random and independent we can rewrite the probability $P({\bs j}|w)$ as
\[{\rm P}\left({\bs j}|w\right)=\prod_{i=1}^{s}{\rm
  P}\left(j_{i}|w\right).\]
The probability ${\rm P}\left(j_{i}|w\right)$
     can be upper-bounded  as
\[{\rm P}\left(j_{i}|w \right)\le \frac{|{\mathcal H}_{i}({\bs
    v},w,j_{i})|}{\binom{ncl}{w}}\]
where ${\mathcal H}_{i}({\bs v},w,j_{i})=\{H_{i}\left|\right. {\bs v}H_{i}^{T}=0,
w,j_{i}\}$. The cardinality of ${\mathcal
H}_{i}({\bs v},w,j_{i})$ can be upper-bounded as
\[|{\mathcal H}_{i}({\bs v},w,j_{i})|=\sum_{w_{k}\ge 1,\sum w_{k}=w}
  \binom{n}{j_{i}}\prod_{k=1}^{j_{i}} \binom{cl}{w_{k}}\]
\begin{equation}
 \le \binom{n}{j_{i}}\binom{cl}{w/j_{i}}^{j_{i}}\binom{w-1}{j_{i}-1}
  \label{50}
\end{equation}
where the sum is upper-bounded  by the maximal term  times the number
of terms $\binom{w-1}{j_{i-1}}$.
 \hfill\QED

Notice that in the above derivations we ignored the fact that
$w/j_{i}$ can be noninteger since we consider the asymptotic behaviour
of (\ref{main}).

It follows from Lemma 1 that
\[{\rm P}({\bs v}H_{\rm wg}^{\rm T}=0|w)\le\sum_{{\bs
    j}}\binom{nlc}{w}^{1-s}\prod_{i=1}^{s}2^{-j_{i}cl(1-R^{c})}\binom{n}{j_{i}}\binom{cl}{w/j_{i}}^{j_{i}}\binom{w-1}{j_{i}-1}\]
\[\le (n+1)^{s}\binom{nlc}{w}^{1-s}\max_{{\bs
    j}}\prod_{i=1}^{s}2^{-j_{i}cl(1-R^{c})}\binom{n}{j_{i}}\binom{cl}{w/j_{i}}^{j_{i}}\binom{w-1}{j_{i}-1}\]
\[=
(n+1)^{s}\binom{nlc}{w}^{1-s}\prod_{i=1}^{s}\max_{j_{i}}2^{-j_{i}cl(1-R^{c})}\binom{n}{j_{i}}\binom{cl}{w/j_{i}}^{j_{i}}\binom{w-1}{j_{i}-1}\]
\begin{equation}
=
(n+1)^{s}\binom{nlc}{w}^{1-s}\left(\max_{j}2^{-jcl(1-R^{c})}\binom{n}{j}\binom{cl}{w/j}^{j}\binom{w-1}{j-1}\right)^{s}.
\label{mainf}
\end{equation}

Consider the asymptotic behaviour of (\ref{main})
 when $m$ tends to infinity. Introduce the notations  $\gamma=j/n$ and
 $\delta=w/(ncl)$ and  the  function
\[F(\delta)=\lim_{l \rightarrow \infty}\frac{\log_{2}{\rm P}({\bs v}H^{\rm
 T}=0|w)}{nlc}.\]
After simple derivations we obtain
\begin{equation}
F(\delta)\le \hat{F}(\delta)\triangleq \left\{
 \max_{\gamma \in (0,1]}
(1-s)h(\delta)-(1-R_{\rm wg})\gamma+s\gamma h\left(\frac{\delta}{\gamma}\right)
\right\}
\label{asympt}
\end{equation}
where $R_{\rm wg}=s(R^{c}-1)+1$ is the  rate of binary woven graph code.
Maximizing (\ref{asympt}) over $0<\gamma\le 1$ gives
\[\gamma_{opt}=\min\left\{1,\frac{\delta}{1-2^{(R_{\rm wg}-1)/s}}\right \}.\]
Inserting $\gamma_{opt}<1$ and $\gamma_{opt}=1$ into (\ref{asympt})
we obtain
\begin{equation}
\hat{F}(\delta)=\left\{
\begin{array}{ll}
h(\delta)+R_{\rm wg}-1 {\mbox,}&{\mbox{ if }} 0<\delta\le 1-2^{(R_{\rm wg}-1)/s}\\
(1-s)h(\delta)-\delta s
\log_{2}\left(2^{-(R_{\rm wg}-1)/s}-1\right){\mbox,}& {\mbox{ if }}\delta \ge
1-2^{(R_{\rm wg}-1)/s}
\end{array}
\right.
\label{f}
\end{equation}
which coincides with (9) and  (10) in  \cite{barg} for $s=2$, that is, if
the graph is bipartite.

For any  $R_{\rm wg}$ and $\delta$  from  $\hat{F}(\delta)<0$, it follows that  there exist
codes of rate $R_{\rm wg}$ with relative minimum distance $\delta_{\rm
  wg}=\delta$. Let
$\delta(R_{\rm wg})$ denote the solution of the equation
\begin{equation} 
\hat{F}(\delta)=0 \label{fun} 
\end{equation}
for $0<\delta\le 1-2^{(R_{\rm wg}-1)/s}$ and let $\delta_{\scriptsize
  VG}(R_{\rm wg})$ be the solution
of $h(\delta)+R_{\rm wg}-1=0$.
Solving (\ref{fun}) for $\gamma_{opt}<1$ and $\gamma_{opt}=1$ we
obtain that there exist woven graph codes of rate $R_{\rm wg}$ with the relative minimum
distance $\delta_{\rm wg}$ satisfying the inequalities:
 \begin{equation}
\delta_{\rm wg}\ge
\left\{
\begin{array}{ll}
\delta(R_{\rm wg})-\epsilon{\mbox,}& {\mbox{ if
  }}R_{\rm wg}>1+s\log_{2}(1-\delta_{{\rm \tiny VG}}(R_{\rm wg}))\\
\delta_{{\rm \tiny VG}}(R_{\rm wg})-\epsilon {\mbox,} &{\mbox{ if
  }}R_{\rm wg}\le 1+s\log_{2}(1-\delta_{{\rm \tiny VG}}(R_{\rm wg})).
\end{array}
\right.
\label{bound}
\end{equation}
\hfill\QED

\begin{figure}[h!]
\center{ \epsfig{file=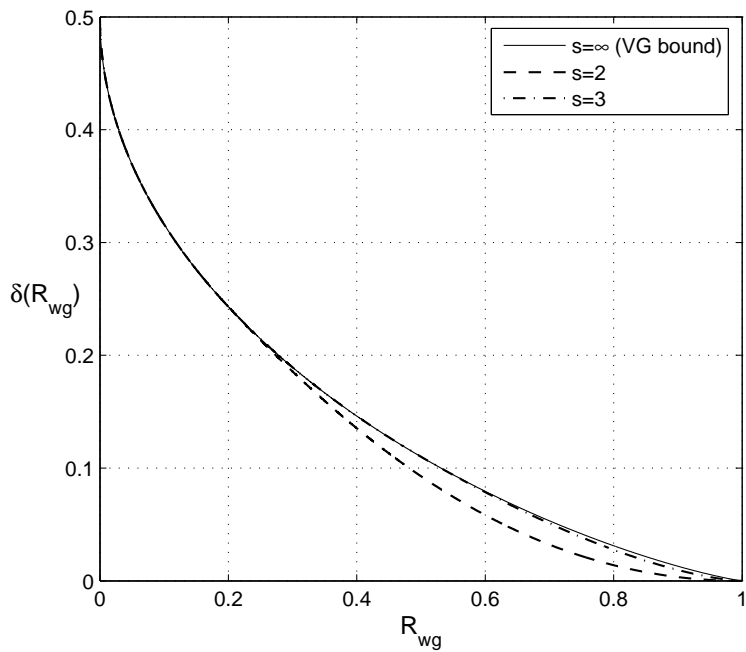,width=110mm}}
\vspace*{-2mm}\caption{The relative minimum distance as a function of
  the code rate for  the ensemble of binary woven graph codes  with block constituent codes.} \label{figure:rd}
\end{figure}
\begin{figure}[h!]
\center{ \epsfig{file=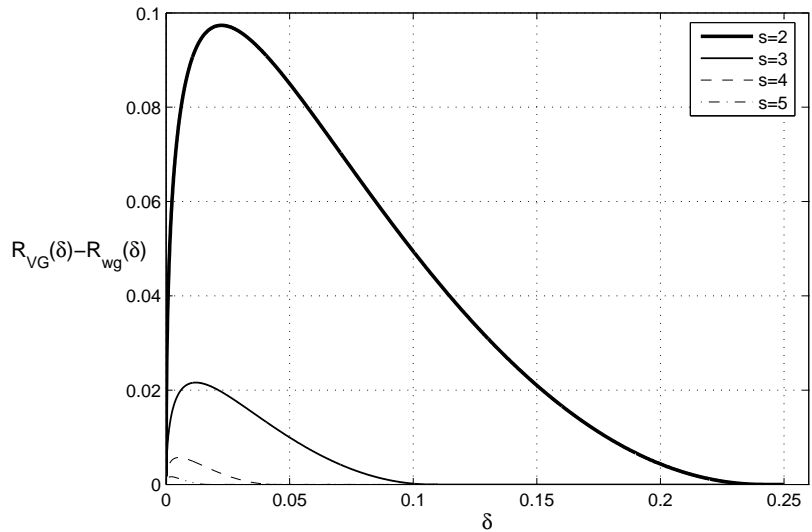,width=110mm}}
\vspace*{-2mm}\caption{The  gap between  the VG bound  and the code
  rate as   a function of the relative minimum distance.} \label{figure:rd2}
\end{figure}
In Fig. \ref{figure:rd} the lower bound (\ref{VGbound}) on the
relative minimum distance for the ensemble of
binary woven graph codes  with block constituent codes as a function of the code rate is shown. It is
easy to see that when $s$ grows the ensemble of
binary woven graph codes contains codes meeting the VG bound for
almost all rates $0\le R_{\rm wg} \le 1$. Fig.  \ref{figure:rd2}  demonstrates
the gap $R_{\rm VG}-R_{\rm wg}$  between the VG bound and
the code rate  as a function of the relative
minimum distance $\delta_{\rm wg}$ for different values of $s$. It follows
from Fig.  \ref{figure:rd2}  that for $s\ge 3$ the difference in code rate
compared to the VG bound is negligible.

\subsection{Asymptotic bound on the free distance of woven graph codes
  with constituent convolutional codes}

Consider a ZT convolutional woven graph code with 
constituent ZT convolutional codes of rate $R^{c}=b/c$.
The length of a ZT woven graph codeword in $nc$-tuples is equal
to $l+m_{\rm wg}$ where $l$ is the number of $nc$-tuples  influenced
by information symbols and $m_{\rm wg}$ is the memory
of the woven graph code of rate $R_{\rm wg}=s(R^{c}-1)+1$.
Denote by $d^{\rm wg}_{{\rm free}}$ the free distance of the corresponding woven graph code.

Now we can prove the following
\vspace*{3mm}
\begin{theorem}(Costello lower bound)
For any $\epsilon > 0$,  some $m_{0}>0$, some integer $s\ge 2$,
and for all $m_{\rm wg}>m_{0}$
in the random ensemble of rate $R_{\rm wg}=s(R^{c}-1)+1$  woven
graph codes over $s$-partite, $s$-uniform, $c$-regular hypergraphs
with constituent convolutional  codes of
rate $R^{c}=b/c$
there exists a code with memory $m_{\rm wg}$
such that its relative free  distance
$\delta^{\rm wg}_{{\rm free}}=d^{\rm wg}_{{\rm free}}/ncm_{\rm wg}$
satisfies the Costello lower bound \cite{RolfKam},
\begin{equation}
\label{CostelloBound}
\delta^{\rm wg}_{{\rm free}}
\ge-\frac{R_{\rm wg}}{\log_{2}\left(2^{1-R_{\rm wg}}-1 \right)}-\epsilon.
\end{equation}
\end{theorem}

{\em Proof}:
Analogously to the derivations in the proof of Theorem \ref{VGtheorem}
let ${\bs j}=(j_{1},j_{2},\dots,j_{s})$ where $j_{i}$ denotes the
number of nonzero constituent codewords in the $i$th subgraph
corresponding to the codeword of weight $w$, $j_{i}\in\{1,...,n\}$.
In order to evaluate the number of nonzero constituent codewords among the $n$
constituent codewords, notice that the set of such codewords is a union of
sets of nonzero constituent codewords belonging to each of the 
$s$ subgraphs. The
cardinality of the union is at least $j_{\max}=\max_i \{j_i\}$.
Therefore the all-zero ``tail'' required to force the encoder into
the zero state has length
at least  $j_{\max} c m_{\rm wg}$.  The total number of redundant
symbols consists of two parts: the number $\sum_{i=1}^s j_icl(1-R^c)$
of parity-check symbols for the nonzero constituent codewords in the $s$
subgraphs and  at least $j_{\max} c m_{\rm wg}$ redundant symbols required
for zero-tail terminating of the woven graph code.
Thus, formula (\ref{redunds}) can be rewritten as

\begin{equation}
{\rm P}({\bs v}H_{\rm wg}^{\mathrm T}={\bs 0}|w,{\bs j})\le
\binom{nc(l+m_{\rm wg})}{w}
\left(
\prod_{i=1}^{s}2^{-j_{i}cl(1-R^{c})}
\right)2^{-j_{\max} c m_{\rm wg}}
.
\label{redunds_conv}
\end{equation}

The statement of the Lemma \ref{Lemma} is changed in a following way
\begin{equation}
P\left({\bs j}|w\right)\le \prod_{i=1}^{s}\frac{\binom{n}{j_{i}}
\binom{c(l+m_{\rm wg})}{w/j_{i}}^{j_{i}}
\binom{w-1}{j_{i}-1}}{\binom{nc(l+m_{\rm wg})}{w}}.
\label{lemf2}
\end{equation}

Instead of (\ref{mainf}) we now have 
\begin{eqnarray}
{\rm P}({\bs v}H_{\rm wg}^{\rm T}=0|w) &\le&
(n+1)^{s}\binom{nc(l+m_{\rm wg})}{w}^{1-s}\nonumber\\
&\times&  
\max_{j}\left\{
\left(2^{-jcl(1-R^{c})}\binom{n}{j}
\binom{c(l+m_{\rm wg})}{w/j}^{j}\binom{w-1}{j-1}\right)^{s}
2^{-jc m_{\rm wg}}
\right\}.
\label{mainf2}
\end{eqnarray}

By introducing the notations
\[
\delta=\frac{w}{ncm_{\rm wg}} \mbox{ , } \mu=\frac{l}{m_{\rm wg}}
\mbox{ , }
\gamma=\frac{j}{n}
\]
we obtain from  (\ref{redunds_conv})--(\ref{mainf2})  that
\begin{eqnarray}
F(\delta)&=&\lim_{m_{\rm wg} \rightarrow \infty} \frac{\log_{2}{\rm
    P}({\bs v}H^{\rm T}=0|w)}{nc m_{\rm wg}}\nonumber\\
&\le& \max_{\gamma\in(0.1]} \left\{
(1-s)(1+\mu)h\left(\frac{\delta}{1+\mu}\right)-\gamma(1+\mu-\mu
R_{\rm wg})+\gamma ( 1+\mu) sh\left(\frac{\delta}{\gamma(1+\mu)}\right)
\right\}.
\label{mainfconv}
\end{eqnarray}
Maximizing (\ref{mainfconv}) over $0<\gamma\le 1$,  we obtain
\begin{equation}
\gamma_{opt}=\min\left\{1,\frac{\delta}{(1+\mu)(1-2^{-x})}\right\}
\label{gam}
\end{equation}
where
\[x=\frac{1+\mu(1-R_{\rm wg})}{s(1+\mu)}.\]

If $s$ is large enough, then $\gamma_{opt}=1$. It follows from
(\ref{mainfconv}) that
\begin{equation}
F(\delta)\le
(1+\mu)h\left(\frac{\delta}{1+\mu}\right)-1-\mu+\mu
R_{\rm wg}{\mbox,}
\label{final}
\end{equation}
Maximization  of  $F(\delta)$  over $\mu$ gives
 \begin{equation}
F_{opt}(\delta) \le -\delta
\log_{2}\left(2^{1-R_{\rm wg}}-1\right)-R_{\rm wg}
\label{cost}
\end{equation}
where
\[\mu_{opt}=\frac{\delta}{1-2^{R_{\rm wg}-1}}-1 .\]
We can find a bound on
$\delta^{\rm wg}_{{\rm free}}$ by  solving  $F_{opt}(\delta)=0$.
Thus, we can conclude that for any $\epsilon>0$ we can find 
a woven graph code such that  (\ref{CostelloBound}) holds
\hfill\QED

\section{Example}

We start with considering a graph code determined by the parity-check
matrix (\ref{incdmatr}).
As  mentioned before,
the matrix (\ref{incdmatr}) can be considered as a $14\times 21$
parity-check matrix. Since the parity checks defined by the graph are
linearly dependent (the sum of the rows of (\ref{incdmatr}) is  equal to
zero) it turned out that by ignoring one parity check we
obtain a parity-check matrix of a   $(21,8)$ linear block
code.  For simplicity we consider the rate $R_{\rm g}=1/3$ code that is obtained by ignoring  the
eighth information symbol which yields  a
$(21,7)$ subcode of this code.

It is easy to see that renumbering the
graph vertices  by adding to each vertex  number some fixed number modulo the
total number of vertices  preserves both the incidence and adjacency
matrices of the graph.
For example, in Fig. \ref{figure:graphH},  by adding 2 modulo 14 we
will get exactly the same graph. When we have a similar property for
linear codes we call such codes quasi-cyclic codes and these block
codes  can
be described as tailbiting (TB) convolutional  codes. Renumbering the  vertices
corresponds to permuting  the rows of (\ref{incdmatr}).
By row permutations,  (\ref{incdmatr})  can be reduced to the
form
\begin{equation}
\footnotesize
\left
(
\begin{array}{ccc:ccc:ccc:ccc:ccc:ccc:ccc}
1&1&1&0&0&0&0&0&0&0&0&0&0&0&0&0&0&0&0&0&0 \\
1&0&0&0&1&0&0&0&0&0&0&1&0&0&0&0&0&0&0&0&0 \\
\hdashline
0&0&0&1&1&1&0&0&0&0&0&0&0&0&0&0&0&0&0&0&0 \\
0&0&0&1&0&0&0&1&0&0&0&0&0&0&1&0&0&0&0&0&0\\
\hdashline
0&0&0&0&0&0&1&1&1&0&0&0&0&0&0&0&0&0&0&0&0\\
0&0&0&0&0&0&1&0&0&0&1&0&0&0&0&0&0&1&0&0&0\\
\hdashline
0&0&0&0&0&0&0&0&0&1&1&1&0&0&0&0&0&0&0&0&0\\
0&0&0&0&0&0&0&0&0&1&0&0&0&1&0&0&0&0&0&0&1\\
\hdashline
0&0&0&0&0&0&0&0&0&0&0&0&1&1&1&0&0&0&0&0&0\\
0&0&1&0&0&0&0&0&0&0&0&0&1&0&0&0&1&0&0&0&0\\
\hdashline
0&0&0&0&0&0&0&0&0&0&0&0&0&0&0&1&1&1&0&0&0\\
0&0&0&0&0&1&0&0&0&0&0&0&0&0&0&1&0&0&0&1&0\\
\hdashline
0&0&0&0&0&0&0&0&0&0&0&0&0&0&0&0&0&0&1&1&1\\
0&1&0&0&0&0&0&0&1&0&0&0&0&0&0&0&0&0&1&0&0
\end{array}
\right
).
\label{modifmatr}
\end{equation}
It follows from  (\ref{modifmatr}) that the graph
shown in Fig. \ref{figure:graphH} corresponds to  a $(21,7)$  TB
code with a parent convolutional code determined by the parity-check matrix
\begin{equation}
 H_{\rm conv}(D)=\left(
\begin{array}{lll}
1&1&1\\
1&D&D^{3}
\end{array}
\right).
\label{convmatr}
\end{equation}
It means that this code is ``tail-bitten'' at the 7th level of the
trellis diagram. A corresponding polynomial generator matrix of the parent
convolutional code has the form
\begin{equation}
 G_{\rm conv}(D)=\left(
\begin{array}{lll}
D+D^{2}&1+D+D^{2}&1
\end{array}
\right).
\label{genmatrc}
\end{equation}

The minimum distance of the $(21,7)$ TB  code is equal to the
graph girth, that is, $d_{\min}=g=6$.

Notice that many regular bipartite graphs look very similar to the Heawood graph in the
sense that by manipulating the incidence (parity-check) matrices and
truncating lengths we can obtain infinite families of
graphs. Some properties of these graphs can be easily predicted from the
properties of the corresponding parent convolutional codes.

Consider the  parity-check matrix (\ref{woven})  of the woven graph code based on
the Heawood bipartite  graph with constituent convolutional codes of rate
$R^{c}=2/3$. This woven graph code has the rate $R_{\rm wg}=2/3\cdot2-1=1/3$.

Let  the rate $R^{c}=2/3$ constituent convolutional code of memory
$m=3$ and overall constraint length
$\nu^{c}=5$ with $d_{\rm free}^{c}=6$ be given by the generator matrix
\begin{equation}
G^{c}(D)=
\begin{pmatrix}
1+D^{2}&D^{2}& 1+D+D^{2}  \\
D+D^{2}+D^{3}& 1&1+D^{2}
\end{pmatrix}. \label{80}
\end{equation}
A corresponding parity-check matrix $H^{c}(D)$ is
\begin{equation}
H^{c}(D)=\begin{pmatrix}
1+D+D^{4}\\
1+D+D^{3}+D^{4}+D^{5}\\
1+D^{2}+D^{3}+D^{4}+D^{5}
\end{pmatrix}^{\mathrm{T}}. \label{90}
\end{equation}
Notice that the constituent code ${\mathcal
  C}^{c}$ considered as a block code over ${\mathds F}((D))$
represents a $(3,2)$ block code with the minimum distance $d_{\rm
  block}^{c}=2$.

By using the  product-type lower bound (\ref{bipartbound}) we
obtain
\[d^{\rm wg}_{{\rm free}}\ge (g/2)d_{\rm free}^{c}=3\times 6=18.\]
On the other hand,  it was verified by computer search that any codeword
of the woven graph code determined by (\ref{woven}) consists of at
least three nonzero codewords of the component code ${\mathcal C}^{c}$
described by (\ref{90}). Moreover, it was found by computer search that each of
these nonzero codewords of ${\mathcal C}^{c}$ has the minimum block
weight $d_{\rm block}^{c}=2$. Note that the codewords of the block
code over ${\mathds F}((D))$ with block weight $d_{\rm block}^{c}=2$
corresponds to the codewords of the convolutional code belonging
to its subcodes of rate $R^{c}=1/2$.
 These three subcodes have generator matrices
\setlength\arraycolsep{2.0 pt}
\[ G_{1}^{c}(D)=
\left(
\begin{array}{cc}
 g_{1}^{c}(D)&g_{2}^{c}(D)
\end{array}
\right)
\]
\[ G_{2}^{c}(D)=
\left(
\begin{array}{cc}
 g_{3}^{c}(D)& g_{2}^{c}(D)
\end{array}
\right)
\]
\[ G_{3}^{c}(D)=
\left(
\begin{array}{cc}
 g_{1}^{c}(D)& g_{3}^{c}(D)
 \end{array}
\right)
\]
where $g_{1}^{c}(D)=1+D+D^{3}+D^{4}+D^{5}$, $g_{2}^{c}(D)=1+D+D^{4}$, and
$g_{3}^{c}(D)=1+D^{2}+D^{3}+D^{4}+D^{5}$.

The minimum free distance over all these
subcodes of rate  $R^{c}=1/2$ is equal to 8. Taking into account that all
other codewords of the woven graph code contain at least four nonzero
codewords of ${\mathcal C}_{c}$ of block weight $d_{\rm block}^{c}=3$
we obtain an improved  lower bound on the free distance of
the woven graph code  as $d_{\mathrm {free}} \ge
\min\{3\times 8,4\times 6\}=24$.

In order to obtain an upper bound on the free distance of the woven graph
code we consider the parity-check matrix (\ref{woven}) in more detail. It
also describes a quasi-cyclic code and can by row permutations be
reduced to a parity-check matrix of a 
two-dimensional code, a TB (block) code in one dimension and a convolutional
code in the other,
 \begin{equation}
H_{\rm wg}(D)=
\footnotesize
\left(
\begin{array}{ccc:ccc:ccc:ccc:ccc:ccc:ccc}
h^{c}_{1}&h^{c}_{2}&h^{c}_{3}&0&0&0&0&0&0&0&0&0&0&0&0&0&0&0&0&0&0\\
t^{c}_{1}&0&0&0&t^{c}_{2}&0&0&0&0&0&0&t_{3}^{c}&0&0&0&0&0&0&0&0&0\\
\hdashline
0&0&0&h_{1}^{c}&h_{2}^{c}&h_{3}^{c}&0&0&0&0&0&0&0&0&0&0&0&0&0&0&0\\
0&0&0&t_{1}^{c}&0&0&0&t_{2}^{c}&0&0&0&0&0&0&t_{3}^{c}&0&0&0&0&0&0\\
\hdashline
0&0&0&0&0&0&h_{1}^{c}&h_{2}^{c}&h_{3}^{c}&0&0&0&0&0&0&0&0&0&0&0&0\\
0&0&0&0&0&0&t_{1}^{c}&0&0&0&t_{2}^{c}&0&0&0&0&0&0&t_{3}^{c}&0&0&0\\
\hdashline
0&0&0&0&0&0&0&0&0&h_{1}^{c}&h_{2}^{c}&h_{3}^{c}&0&0&0&0&0&0&0&0&0\\
0&0&0&0&0&0&0&0&0&t_{1}^{c}&0&0&0&t_{2}^{c}&0&0&0&0&0&0&t_{3}^{c}\\
\hdashline
0&0&0&0&0&0&0&0&0&0&0&0&h_{1}^{c}&h_{2}^{c}&h_{3}^{c}&0&0&0&0&0&0\\
0&0&t_{3}^{c}&0&0&0&0&0&0&0&0&0&t_{1}^{c}&0&0&0&t_{2}^{c}&0&0&0&0\\
\hdashline
0&0&0&0&0&0&0&0&0&0&0&0&0&0&0&h_{1}^{c}&h_{2}^{c}&h_{3}^{c}&0&0&0\\
0&0&0&0&0&t_{2}^{c}&0&0&0&0&0&0&0&0&0&t_{3}^{c}&0&0&0&t_{1}^{c}&0\\
\hdashline
0&0&0&0&0&0&0&0&0&0&0&0&0&0&0&0&0&0&h_{1}^{c}&h_{2}^{c}&h_{3}^{c}\\
0&t_{1}^{c}&0&0&0&0&0&0&t_{2}^{c}&0&0&0&0&0&0&0&0&0&t_{3}^{c}&0&0
\end{array}
\right)
\label{modwoven}
\end{equation}
A parity-check matrix of the parent convolutional code for the  TB
code (\ref{modwoven}) given in  symbolic form is
\setlength\arraycolsep{5.0 pt}
   \begin{equation}
H(D,Z)= \left(
\begin{array}{lll}
h^{c}_{1}(D)&h^{c}_{2}(D)&h^{c}_{3}(D)  \\
t^{c}_{1}(D)&t^{c}_{2}(D)Z&t^{c}_{3}(D)Z^{3}
\end{array}
 \right)
\label{97}
\end{equation}
where $Z$ and $D$ are formal variables. The matrix (\ref{97})
can be considered as a parity-check matrix of a two-dimensional
convolutional code. The variable $Z$ corresponds to the parent
convolutional code of the Heawood graph code (\ref{convmatr}), the
variable $D$ is used for the constituent convolutional code
(\ref{90}).

A generator matrix of the two-dimensional  convolutional code with the
parity-check matrix (\ref{97}) has the form
\setlength\arraycolsep{5.0 pt}
   \begin{equation}
G(D,Z)= \left(
\begin{array}{lll}
g^{c}_{e}(D)Z+g^{c}_{c}(D)Z^{3}&g^{c}_{a}(D)+g^{c}_{d}(D)Z^{3}&g^{c}_{b}(D)+g^{c}_{f}(D)Z \\
\end{array}
 \right)
\label{97gen}
\end{equation}
where $g^{c}_{a}(D)=h^{c}_{3}(D)t^{c}_{1}(D)$, %
      $g^{c}_{b}(D)=h^{c}_{2}(D)t^{c}_{1}(D)$, %
      $g^{c}_{c}(D)=h^{c}_{2}(D)t^{c}_{3}(D)$,       %
      $g_{d}^{c}(D)=h^{c}_{1}(D)t^{c}_{3}(D)$, %
      $g^{c}_{e}(D)=h^{c}_{3}(D)t^{c}_{2}(D)$,  and %
      $g_{f}^{c}(D)=h^{c}_{1}(D)t^{c}_{2}(D)$.

The generator matrix  (\ref{97gen}) tail-bitten over variable $Z$ at
length $21$ yields the generator matrix $G(D)$ of the code (\ref{modwoven}),
\setlength\arraycolsep{2.0pt}
\begin{equation}
\footnotesize
G_{\rm wg}(D)=
\left(
\begin{array}{ccc:ccc:ccc:ccc:ccc:ccc:ccc}
g^{c}_{c}&g^{c}_{d}&0&0&0&0&g^{c}_{e}&0&
g^{c}_{f}&0&g^{c}_{a}&g^{c}_{b}&0&0&0&0&0&0&0&0&0 \\
\hdashline
0&0&0&g^{c}_{c}&g^{c}_{d}&0&0&0&0&g^{c}_{e}&0& g^{c}_{f}&
0&g^{c}_{a}&g^{c}_{b}&0&0&0&0&0&0\\
\hdashline
0&0&0&0&0&0&g^{c}_{c}&g^{c}_{d}&0& 0&0&0&g^{c}_{e}&
0& g^{c}_{f}&0&g^{c}_{a}&g^{c}_{b}&0&0&0\\
\hdashline
0&0&0&0&0&0&0&0&0&g^{c}_{c}&g^{c}_{d}&0&0&0&0&g^{c}_{e}&
0& g^{c}_{f}&0&g^{c}_{a}&g^{c}_{b}\\
\hdashline
0&g^{c}_{a}&g^{c}_{b}&0&0&0&0&0&0&0&0&0&g^{c}_{c}&
g^{c}_{d}&0&0&0&0&g^{c}_{e}&0&g^{c}_{f}\\
\hdashline
g^{c}_{e}&0&g^{c}_{f}&0&g^{c}_{a}&g^{c}_{b}&
0&0&0&0&0&0&0&0&0&g^{c}_{c}&g^{c}_{d}&0& 0&0&0\\
\hdashline
0&0&0&g^{c}_{e}&0&g^{c}_{f}&0&g^{c}_{a}& g^{c}_{b}&0&0&0&0&0&0&0&0&0&g^{c}_{c}&g^{c}_{d}&0
\end{array} \right)
\label{tb13}
\end{equation}
where $g_i^c$ is short-hand for $g_i^c(D)$.

Notice that any of the six permutations of the columns $h^{c}_{i}(D)$,
$i=1,2,3,$  generates  a woven graph code. The permutation $t^{c}_{1}(D)=h^{c}_{1}(D)$,
$t^{c}_{2}(D)=h^{c}_{3}(D)$, and $t^{c}_{3}(D)=h^{c}_{2}(D)$ describes the woven
graph code with the largest free distance.
The overall constraint length of this generator matrix is
  equal to 70 but the matrix is not in minimal form. A  minimal-basic
  generator matrix \cite{RolfKam} has the  overall constraint length equal to 64 and
  differs from (\ref{tb13}) by one row which can replace any of the rows of
  $G(D)$ and has the form
  \setlength\arraycolsep{2.0 pt}
\[
\left(
\begin{array}{ccccccc}
G_{0}(D)&G_{0}(D)&G_{0}(D)&G_{0}(D)&G_{0}(D)&G_{0}(D)&G_{0}(D)
\end{array}
\right)
\]
where
\setlength\arraycolsep{2.0 pt}
\[
 G_{0}(D)=
\left(
\begin{array}{ccc}
g^{c}_{p}(D)&g^{c}_{q}(D)&g^{c}_{q}(D)
\end{array}
\right)
\]
where
$g^{c}_{p}(D)=D+D^{2}$  and  $g^{c}_{q}(D)=1+D+D^{4}$.

The matrix (\ref{tb13})
is a generator matrix of a
convolutional code of rate $R_{\rm wg}=7/21$. By applying the BEAST algorithm
\cite{beast} to  the minimal-basic generator matrices corresponding to
the different permutations of  the columns $h^{c}_{i}(D)$, $i=1,2,3,$
we obtained the  free distance and   a few spectrum coefficients
of the corresponding woven graph codes. The parameters of the best obtained woven graph
codes are presented in Table 1.

\begin{table}[htb]
\caption{Spectra and overall constraint lengths of rate $R_{\rm wg}=1/3$ woven graph codes }
\begin{center}
\scalebox{1.2}{
{\tt
\begin{tabular}{|c|c|c|c|}
\hline $\rm {Permutation}$
& $\nu$&$d_{{\rm free}}$&$\rm {Spectrum}$ \\  \hline\hline
$h^{c}_{1}(D),h^{c}_{3}(D),h^{c}_{2}(D)$& $64$&$32$&$7,0,0,0,0,0,7,0,7,0\dots$  \\\hline
$h^{c}_{2}(D),h^{c}_{1}(D),h^{c}_{3}(D)$ & $65$&$32$&$7,0,0,0,7,0,0,0,21,0\dots$  \\ \hline
$h^{c}_{2}(D),h^{c}_{3}(D),h^{c}_{1}(D)$& $66$&$30$&$7,0,0,0,0,0,0,0,7,0\dots$\\ \hline
\end{tabular}
}}
\end{center}
\end{table}

\section{ Encoding}
Generally speaking,  encoding of  graph-based block codes has complexity
$O(N^{2})$,  where $N$ is the blocklength. This technique implies that we
find a generator matrix corresponding to the given parity-check matrix
and then multiply the information sequence by the obtained generator
matrix. However, we  showed  by examples that some  regular  graph codes
as well as woven graph codes  are
quasi-cyclic codes and thereby they can be interpreted as TB
codes. For this
class of codes the complexity of the encoding is proportional to the
constraint length of the parent convolutional code.

In this section we are going to illustrate  by an example an encoder of a
woven graph code with constituent convolutional codes having encoding
complexity proportional to the overall constraint length of the
corresponding woven graph code $\nu_{\rm wg} \le ns\nu^{c}$.

\begin{figure}[h!]
\center{ \epsfig{file=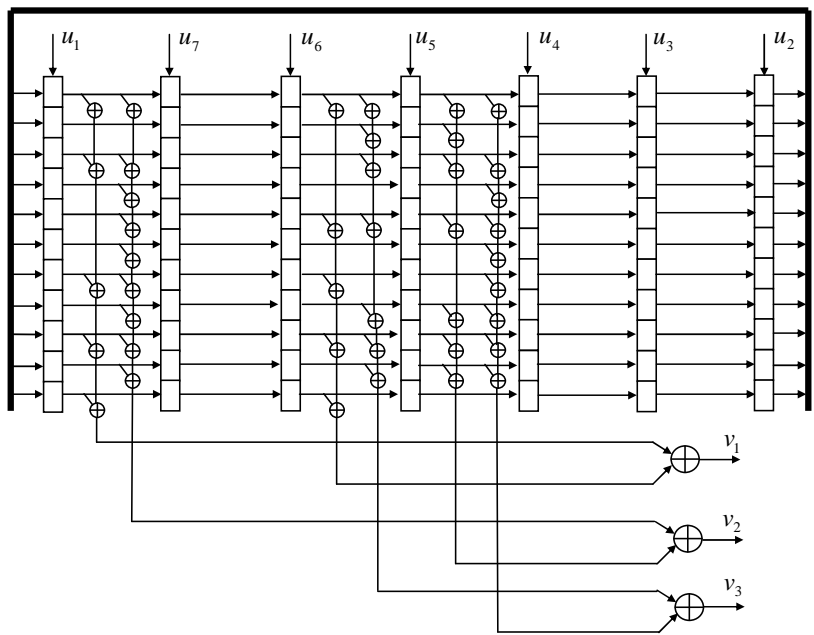,width=110mm}}
\vspace*{-2mm}\caption{An encoder of the two-dimensional woven graph code.}
\label{figure:encoder}
\end{figure}

Consider again the woven graph code in our example. It is based on the
Heawood graph and uses constituent convolutional codes of rate
$R^{c}=2/3$ and  overall constraint length $\nu^{c}=5$.
Taking into account the representation (\ref{tb13}) of the woven graph code
as a rate $R_{\rm wg}=7/21$ two-dimensional code, a TB (block) code in one dimension
and a convolutional code in the other, we can draw its encoder as shown in
Fig. \ref{figure:encoder}.

The input symbols $u_{1},u_{2},\dots,u_{7}$ enter the encoder once per
each cycle of duration  seven time instants. At each time
moment the contents of each  register is rewritten into the next (modulo
seven)  register and the three output symbols $v_{1},v_{2},v_{3}$ are
generated. In other words,  each of  the registers corresponding to the
constituent code can be considered as an  enlarged delay element of the
encoder of the ``TB-dimension'' code determined by the graph. The sequence
$u_{1},u_{2},\dots,u_{7}$ determines a transition between the states of
this encoder. After a
cycle of seven time instants we return to the starting state of the enlarged encoder
and a TB-codeword (or a word from one of its cosets) of length 21 has been  generated. Then the following seven input
symbols $u_{8},u_{9},\dots,u_{14}$ enter and after seven time instants another
word of length 21 has been generated, etc.

\section{Conclusion}

The asymptotic behavior of the woven graph codes with block as well
as with convolutional constituent codes has been studied. It was shown
that in the random ensemble of such codes based on $s$-partite,
$s$-uniform, $c$-regular hypergraphs  we can find a  value $s\ge 2$
such that for any code rate
there exist codes meeting the VG and the Costello lower bound on the
minimum distance and free distance, respectively.
Product-type lower bounds on the
minimum distance of graph-based and woven graph codes have been derived.
Example of a rate $R_{\rm wg}=1/3$  woven graph code  with free
distance above the product bound is presented.
It is shown, by an example, that  woven graph codes can be encoded
with a complexity proportional to the constraint length of the
constituent convolutional code.

\vspace*{2mm}

\section*{Acknowledgment}

This work was supported in part by the Royal Swedish Academy of
Sciences in cooperation with the Russian Academy of Sciences and
in part by the Swedish Research Council under Grant 621-2007-6281.

\end{document}